\documentclass[aip,jcp,reprint,superscriptaddress]{revtex4-2}

\usepackage{graphicx}

\usepackage{amsmath,amssymb}

\usepackage{newtxtext,newtxmath}

\usepackage{bm}

\usepackage{booktabs}

\usepackage{xcolor}

\usepackage{lipsum}

\usepackage[
 colorlinks=true,
 linkcolor=blue,
 citecolor=blue,
 urlcolor=blue
]{hyperref}

\graphicspath{{./}{./fig/}{../fig/}}

\begin{document}
\title{Effects of chain stiffness on the breakdown of the Cox--Merz rule in linear and ring polymer melts
}

\author{Keishin Tsujino}
\affiliation{Division of Chemical Engineering, Department of Materials Engineering Science, Graduate School of Engineering Science, The University of Osaka, Toyonaka, Osaka 560-8531, Japan}

\author{Yuhi Sakamaki}
\affiliation{Division of Chemical Engineering, Department of Materials Engineering Science, Graduate School of Engineering Science, The University of Osaka, Toyonaka, Osaka 560-8531, Japan}

\author{Shota Goto}
\affiliation{Division of Chemical Engineering, Department of Materials Engineering Science, Graduate School of Engineering Science, The University of Osaka, Toyonaka, Osaka 560-8531, Japan}

\author{Kang Kim}
\email{kk@cheng.es.osaka-u.ac.jp}
\affiliation{Division of Chemical Engineering, Department of Materials Engineering Science, Graduate School of Engineering Science, The University of Osaka, Toyonaka, Osaka 560-8531, Japan}

\author{Nobuyuki Matubayasi}
\email{nobuyuki@cheng.es.osaka-u.ac.jp}
\affiliation{Division of Chemical Engineering, Department of Materials Engineering Science, Graduate School of Engineering Science, The University of Osaka, Toyonaka, Osaka 560-8531, Japan}

%%%%%%%%%%%%%%%%%%%%%%%%%%%%%%%%%%%%%%%%%%%%%%%%%%%%%%%%%%%%%%%%%%%%%%%%%%%%%%%%%%%%%%%%%%%%%%%%%%%%%%%%%%%%%%%%%%%%%
\date{\today}

%%%%%%%%%%%%%%%%%%%%%%%%%%%%%%%%%%%%%%%%%%%%%%%%%%%%%%%%%%%%%%%%%%%%%%%%%%%%%%%%%%%%%%%%%%%%%%%%%%%%%%%%%%%%%%%%%%%%%
\begin{abstract}
The Cox--Merz rule is an empirical relation between the magnitude of the
 complex viscosity, $\lvert\eta^*(\omega)\rvert$, obtained from linear
 viscoelasticity, and the steady-shear viscosity, $\eta(\dot{\gamma})$. 
In this study, we performed molecular dynamics simulations of
 coarse-grained polymer melts using the Kremer--Grest model and
 systematically examined the validity of the Cox--Merz rule as a
 function of chain length and chain stiffness for linear and ring
 polymers. 
For short chains and, more generally, for flexible chains,
 $\lvert\eta^*(\omega)\rvert$ and $\eta(\dot{\gamma})$ show good
 agreement at corresponding values of angular frequency $\omega$ and
 shear rate $\dot{\gamma}$. 
For linear polymers, however, the discrepancy between the two
 viscosities becomes increasingly pronounced with increasing chain
 length and chain stiffness, with $\lvert\eta^*(\omega)\rvert$ exceeding
 $\eta(\dot{\gamma})$ at high corresponding values of $\omega$ and
 $\dot{\gamma}$. 
This deviation is associated with the pronounced nonlinear response
 under strong steady shear, where flow-induced chain extension and
 alignment can modify the entanglement constraints that govern
 relaxation near equilibrium. 
In contrast, ring polymers retain substantially better agreement between
 $\lvert\eta^*(\omega)\rvert$ and $\eta(\dot{\gamma})$ with increasing
 chain length and stiffness within the range examined here, indicating a
 weaker breakdown of the Cox--Merz rule.
These results demonstrate that the applicability of the Cox--Merz rule
 depends strongly on chain length, chain stiffness, and molecular
 architecture. 
When the molecular response under steady shear deviates substantially
 from the relaxation behavior characterized by linear viscoelasticity
 near equilibrium, the complex viscosity can no longer accurately
 predict the steady-shear viscosity, resulting in a pronounced breakdown
 of the Cox--Merz rule.
\end{abstract}
%%%%%%%%%%%%%%%%%%%%%%%%%%%%%%%%%%%%%%%%%%%%%%%%%%%%%%%%%%%%%%%%%%%%%%%%%%%%%%%%%%%%%%%%%%%%%%%%%%%%%%%%%%%%%%%%%%%%%
\maketitle
%%%%%%%%%%%%%%%%%%%%%%%%%%%%%%%%%%%%%%%%%%%%%%%%%%%%%%%%%%%%%%%%%%%%%%%%%%%%%%%%%%%%%%%%%%%%%%%%%%%%%%%%%%%%%%%%%%%%%

\section{Introduction}

The viscoelastic response of polymeric materials is governed by
molecular relaxation processes occurring over a broad range of time
scales. 
A central quantity characterizing these processes is the stress
relaxation modulus, $G(t)$, which describes the decay of stress
following a small step deformation and provides a direct connection
between molecular dynamics and macroscopic linear
viscoelasticity.~\cite{doi1986Theory, rubinstein2003Polymer}
The characteristic form
and time scales of $G(t)$
reflect the underlying chain dynamics, including Rouse relaxation and,
for sufficiently long linear polymers, relaxation associated with
entanglement constraints. 
In the tube picture, these constraints confine the motion of an
entangled linear chain to an effective tube formed by the surrounding
chains, leading to reptation as a characteristic long-time relaxation
mechanism.~\cite{degennes1971Reptation}
In the entangled regime, $G(t)$ exhibits a plateau characterized by the
plateau modulus, $G_N^0$, which is directly related to the entanglement
length $N_\mathrm{e}$; a smaller $N_\mathrm{e}$ corresponds to a higher
density of entanglement constraints and hence a larger plateau modulus.
The frequency-dependent storage and loss moduli, $G'(\omega)$ and
$G''(\omega)$, can be obtained from $G(t)$, and together determine the
complex viscosity, $\lvert\eta^*(\omega)\rvert$. 
Thus, $G(t)$ provides a basis for relating molecular relaxation
processes and entanglement constraints to the frequency-dependent
rheological response of polymer melts.

The Cox--Merz rule is an empirical relation between the linear
viscoelastic response measured under oscillatory shear and the nonlinear
viscosity measured under steady shear.~\cite{cox1958Correlation}
Specifically, the magnitude of the complex viscosity,
$\lvert\eta^*(\omega)\rvert$, measured as a function of angular
frequency $\omega$, is compared with the steady-shear viscosity,
$\eta(\dot{\gamma})$, measured as a function of shear rate
$\dot{\gamma}$. 
The rule states that these two quantities approximately coincide when
evaluated at the same numerical value of their respective arguments,
\begin{equation}
\lvert\eta^*(\omega)\rvert_{\omega=\dot{\gamma}}
\simeq
\eta(\dot{\gamma}).
\end{equation}
Both $\omega$ and $\dot{\gamma}$ have dimensions of inverse time, but
they characterize different deformation protocols: $\omega$ sets the
time scale of an oscillatory deformation, whereas $\dot{\gamma}$
specifies the rate of strain accumulation under steady
shear. 
Accordingly, the condition $\omega=\dot{\gamma}$ should be
understood as an empirical identification of these inverse time scales,
expressed in the same units, rather than as a physical equivalence
between angular frequency and shear rate. 
Although this correspondence
has been observed for a wide range of polymer melts and solutions, it
has no general theoretical basis because $\lvert\eta^*(\omega)\rvert$
characterizes the linear response near equilibrium, whereas
$\eta(\dot{\gamma})$ reflects the nonlinear response under steady
flow. 
Deviations from the Cox--Merz rule may therefore reflect
differences between equilibrium relaxation processes and the structural
or conformational response induced by steady shear.

Parisi \textit{et al.} combined experiments, molecular dynamics (MD)
simulations using the Kremer--Grest model,~\cite{kremer1990Dynamics} and
scaling theory to examine the steady-shear rheology of ring polymer
melts over a wide range of chain lengths, with the simulations extending
up to $N=800$.~\cite{parisi2021Nonlinear} 
The study showed that ring polymers exhibit weaker shear thinning than
their linear counterparts and that the shear viscosity follows an
approximately chain-length-independent power law over an intermediate
range of shear rates. 
At higher shear rates, however, the viscosity becomes dependent on chain
length. 
These behaviors were interpreted using a shear-slit model in which the
conformations and dynamics of ring polymers under shear are governed by
two characteristic length scales: the tension-blob size and the larger
shear-slit thickness associated with confinement in the
velocity-gradient direction. 
The same framework was used to rationalize the approximate validity of
the Cox--Merz rule for ring polymers. 
However, the simulations in that study were restricted to a single chain
stiffness, $\varepsilon_\theta=1.5$, where $\varepsilon_\theta$ is the
bending potential parameter defined in Eq.~\eqref{eq:bond}.

Our previous studies have shown that molecular architecture and chain
stiffness strongly affect the dynamics of linear and ring polymer melts. 
For the same polymer model at $\varepsilon_\theta=1.5$, we found
different chain-length dependences of the relaxation dynamics for the
two architectures, with a crossover from Rouse to reptation behavior for
linear polymers and Rouse-like behavior for ring
polymers.~\cite{goto2021Effects}
We subsequently examined ring polymer melts at $\varepsilon_\theta=1.5$
and 5 and found that chain stiffness strongly affects their
conformations and dynamics.~\cite{goto2023Unraveling} 
These results suggest that both molecular architecture and chain
stiffness should be considered when examining the relation between
molecular relaxation and rheological response.

In the present study, rather than extending the chain length to larger
$N$, we focus on shorter chains with $10 \leq N \leq 100$ and
systematically vary the chain stiffness over $0 \le \varepsilon_\theta
\le 3$. 
The chain stiffness is expected to affect topological constraints
differently in linear and ring polymers. 
In linear polymer melts,
increasing chain stiffness generally decreases the entanglement length
$N_\mathrm{e}$ and thereby modifies the importance of entanglement
effects at a given chain length. 
The dependence of $N_{\mathrm e}$ on
chain stiffness for the polymer model considered here is discussed in
Sec.~\ref{section_methods}. 
For ring polymers, however, the conventional entanglement
length $N_\mathrm{e}$ defined for linear polymers is not directly
applicable because of their closed-chain architecture and the absence
of chain ends~\cite{halverson2011Molecular, halverson2011Moleculara,
halverson2012Rheology, ubertini2022Entanglement};
instead, their dynamics are influenced by architecture-specific
topological constraints, including inter-ring
threadings~\cite{michieletto2014Threading, lee2015Slowing,
michieletto2016Topologically, tsalikis2016Analysis, michieletto2017Ring,
smrek2019Threading}.
Increasing stiffness suppresses local bending and favors more extended
and open ring conformations, which may facilitate inter-ring
threadings. 
Despite extensive studies of ring polymer dynamics,
frequency-dependent moduli $G'(\omega)$ and $G''(\omega)$ have been
reported for only selected Kremer--Grest ring polymer
systems.~\cite{murashima2021Viscosity}
Varying both chain length and stiffness thus provides a means of
examining how entanglements in linear polymers and threadings in ring
polymers are related to the Cox--Merz correspondence. 
By comparing the
two architectures under the same conditions, we aim to clarify the
roles of these distinct topological constraints in the validity and
breakdown of the Cox--Merz rule.

\section{Simulation Details}
\label{section_methods}

MD simulations of both linear and ring polymers were performed using the
Kremer--Grest model.~\cite{kremer1990Dynamics}
Each polymer chain consists of $N$ monomer beads with mass $m$ and
diameter $\sigma$. 
The system contains $M$ polymer chains in a three-dimensional cubic
simulation cell of volume $V$, with periodic boundary conditions imposed
in all directions. 
In this study, the chain lengths of the polymers were set to $N = 10$,
20, 50, and 100, and the number of chains was set to $M = 100$.
Non-bonded interactions between beads were described by the
Lennard–Jones (LJ) potential truncated at its minimum, such that only
the repulsive part of the interaction is retained, as given by the
following equation:
\begin{equation}
U_\mathrm{LJ} (r) = 
\begin{cases}
4\varepsilon\left[
\left(\dfrac{\sigma}{r}\right)^{12}
-\left(\dfrac{\sigma}{r}\right)^6
\right]+\varepsilon,
& r\le 2^{1/6}\sigma,\\
0, & r>2^{1/6}\sigma,
\end{cases}
\end{equation}
where $r$ denotes the interbead distance, and 
the LJ potential is shifted to zero at the cutoff distance $r = 2^{1/6}\sigma$.
In addition, adjacent monomer beads along each polymer chain interact
through the finitely extensible nonlinear elastic (FENE) bond potential,
\begin{equation}
U_\mathrm{FENE} (r) = 
\begin{cases}
- \frac{1}{2} K R_0^2 \ln \left[1 - \left(\dfrac{r}{R_0}\right)^2\right],
 & r \le R_0,\\
\infty, & r>R_0,
\end{cases}
\end{equation}
where $K$ is the spring constant and $R_0$ is the maximum
bond extension.
Here, we set $K = 30 \varepsilon/\sigma^2$ and $R_0=1.5 \sigma$.
Chain stiffness was introduced through a bending potential that depends
on the angle $\theta$ between two consecutive bonds:
\begin{equation}
U_\mathrm{bend} (\theta) = \varepsilon_\theta [1 - \cos(\theta -
 \theta_0)].
\label{eq:bond}
\end{equation}
The equilibrium bond angle was set to $\theta_0=180^\circ$.
All simulations were performed using the Large-scale Atomic/Molecular
Massively Parallel Simulator (LAMMPS).~\cite{plimpton1995Fast}
Hereinafter, all physical quantities are expressed in reduced LJ units,
with $\sigma$, $\varepsilon$, and $m$ used as the reference units for length,
energy, and mass, respectively.
The corresponding unit of time is $\tau= \sigma\sqrt{m/\varepsilon}$, and 
the temperature is expressed in units of
$\varepsilon/k_{\mathrm{B}}$, where $k_{\mathrm{B}}$ is
the Boltzmann constant.

%%%%%%%%% Fig. 1 %%%%%%%%%%%%%%%%%%%%%%%%%%%%%%%%%%%%%%%%%%%%%%%%%%%%%%%%%%%%%%%%%%%%%%%%%%%%%%%%%%%%%%%%%%%%%%%%%%%%
\begin{figure}[t]
\centering
 \includegraphics[width=0.45\textwidth]{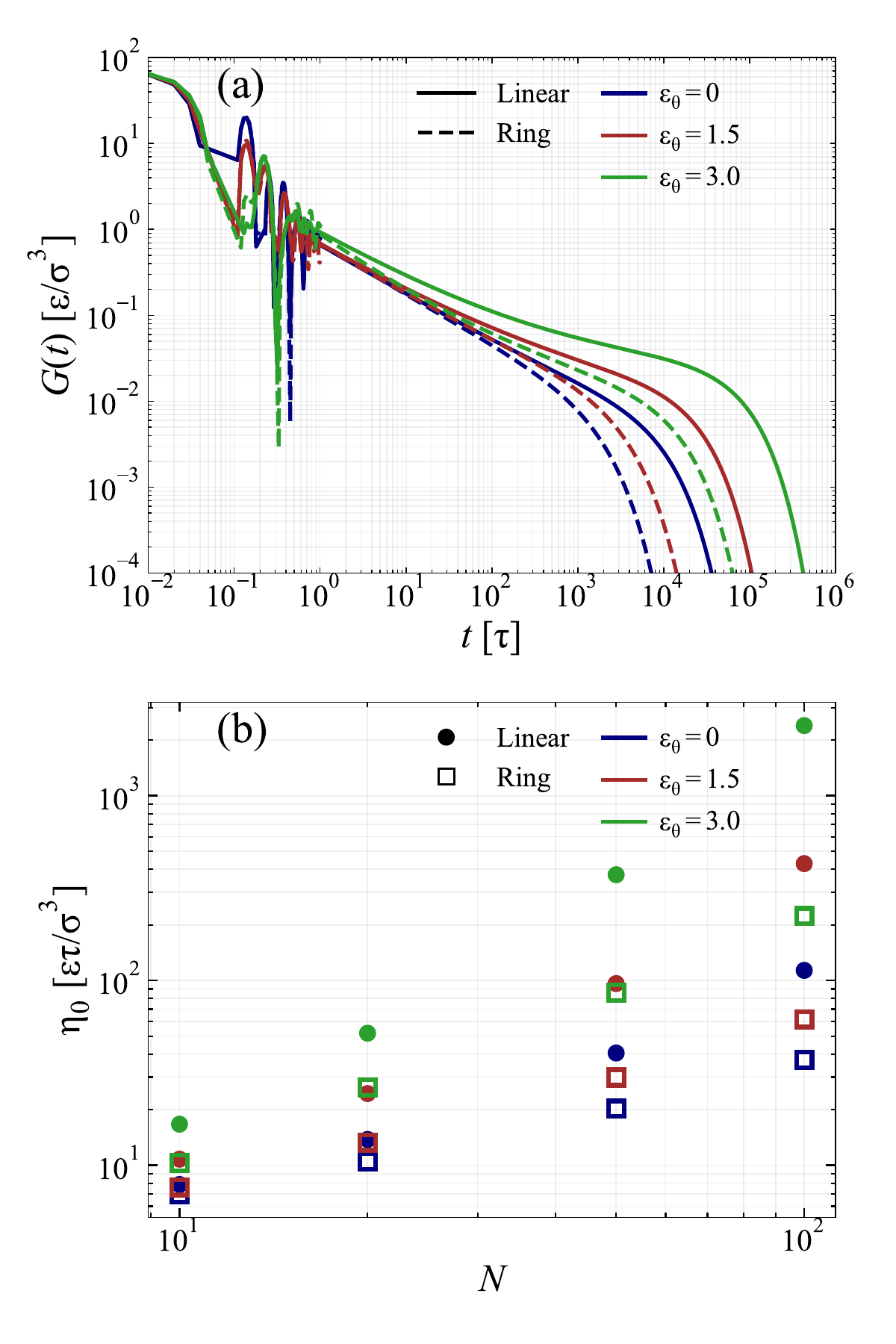}
\caption{(a) Stress relaxation modulus $G(t)$ for linear and ring
 polymers with $N=100$ at different chain stiffnesses $\varepsilon_\theta$. 
Solid and dashed lines represent linear and ring polymers,
 respectively, with $\varepsilon_\theta = 0$ (blue), 1.5 (red), and 3.0
 (green).
(b) Dependence of the 
zero-shear viscosity, $\eta_0$, on chain length $N$.
Circles and squares represent linear and ring polymers, respectively,
 with $\varepsilon_\theta = 0$ (blue), 1.5 (red), and 3.0 (green).
}
\label{fig:stress_relaxation}
\end{figure}
%%%%%%%%%%%%%%%%%%%%%%%%%%%%%%%%%%%%%%%%%%%%%%%%%%%%%%%%%%%%%%%%%%%%%%%%%%%%%%%%%%%%%%%%%%%%%%%%%%%%%%%%%%%%%%%%%%%%%

We performed two types of MD simulations: equilibrium
molecular dynamics (EMD) and nonequilibrium molecular dynamics (NEMD)
simulations.
For each chain length and polymer architecture (linear and ring), the system
was first equilibrated at a temperature of
$T = 1$ and a segment number density of $\rho =
MN / V = 0.85$.
Although $\varepsilon_\theta=1.5$
is commonly employed,~\cite{auhl2003Equilibration, everaers2004Rheology,
hsu2016Static} we also considered
$\varepsilon_\theta=0$ and 3 to examine the effect of chain stiffness.
For linear polymer melts, 
the entanglement length was estimated to be
$N_{\mathrm{e}} \approx 28$ beads for $\varepsilon_\theta =
1.5$.~\cite{everaers2004Rheology, hsu2016Static}
Subsequent studies systematically examined the dependence of
$N_\mathrm{e}$ on chain stiffness.
This dependence was characterized over $0 \le \varepsilon_\theta
\le 2$, yielding $N_{\mathrm e} \approx 86$ for fully flexible chains
$(\varepsilon_\theta = 0)$,~\cite{moreira2015Direct} and was later
examined over an extended range of $-1 \le \varepsilon_\theta \le
2.5$.~\cite{svaneborg2020Characteristic} 
More recent work further extended the range to $0 \le \varepsilon_\theta
\le 5.5$, reporting $N_{\mathrm e} \approx 13$ at
$\varepsilon_\theta=3$.~\cite{dietz2022Validation}
Thus, the three values of $\varepsilon_\theta$ considered here
correspond to substantially different entanglement lengths in linear
polymer melts. 
The range of chain stiffness examined here is also relevant to
experimental polymer melts. 
Everaers \textit{et al.}~\cite{everaers2020Kremer}
mapped the Kremer--Grest model onto a variety of commodity polymers
based on the scaling properties of their Kuhn lengths
and obtained stiffness parameters
in the range $-0.4\leq\varepsilon_\theta\leq2.3$, which substantially
overlaps with the range $0\leq\varepsilon_\theta\leq3$ considered here.
Note that the effects of chain stiffness on the structure and dynamics
of ring polymer melts have also been extensively
studied.~\cite{roy2022Effect, goto2023Unraveling, datta2023Viscosity, ghosh2024Onset}

In the EMD simulations, time-dependent off-diagonal components of the
pressure tensor were computed.
The $\alpha\beta$ component of the pressure tensor,
$P_{\alpha\beta}$, is given by 
\begin{equation}
P_{\alpha\beta}  = \frac{1}{V}\left[ \sum_{i=1}^{N_\mathrm{tot}} m v_{i,\alpha}
 v_{i,\beta}
+W_{\alpha\beta}\right],
%+ \frac{1}{2} \sum_{i=1}^{N_\mathrm{tot}}\sum_{j=1, j\ne
%i}^{N_\mathrm{tot}} r_{ij,\alpha} f_{ij,\beta}\right],
\end{equation}
where $N_\mathrm{tot}=MN$ is the total number of beads, 
$v_{i,\alpha}$ is the $\alpha$ component of the velocity of bead $i$, and
$W_{\alpha\beta}$ denotes the $\alpha\beta$ component of the virial contribution from all interactions.
%$r_{ij,\alpha}$ and $f_{ij,\beta}$ are the $\alpha$ component of the
%relative position vector $\mathbf{r}_{ij} = \mathbf{r}_i -\mathbf{r}_j$,
%and the $\beta$ component of the force between $i$ and $j$, respectively.
Here, $\alpha$ and $\beta$ are $x$, $y$, or $z$.
The autocorrelation functions of off-diagonal components of the pressure
tensor, $\langle P_{\alpha\beta}(t) P_{\alpha\beta}(0)
\rangle$, were calculated.
The corresponding components of the stress relaxation modulus,
$G_{\alpha\beta}(t)$, were then obtained from the
Green--Kubo relation,~\cite{zhou2006Direct, likhtman2007Linear,
lee2009Entangled, hou2010Stress, hsu2016Static, murashima2021Viscosity,
adeyemi2022Equilibrium, tu2023Unexpected,
behbahani2024Relaxation, behbahani2026Stress, yoshimoto2026Effect}
\begin{equation}
G_{\alpha\beta}(t) = \frac{V}{k_\mathrm{B}T} \langle
 P_{\alpha\beta}(t) P_{\alpha\beta}(0)
\rangle.
\end{equation}
The stress relaxation modulus $G(t)$ was obtained by averaging
$G_{\alpha\beta}(t)$ over the three off-diagonal components, 
$xy$, $xz$, and $yz$.
To reduce statistical fluctuations in the stress relaxation modulus
$G(t)$, we performed a least-squares fit 
using the interpolation formulas for linear and ring polymers reported
by Parisi \textit{et al.}~\cite{parisi2021Nonlinear}.
The zero-shear viscosity, $\eta_0$, was obtained by integrating the
stress relaxation modulus $G(t)$ over time:
\begin{equation}
\eta_0  = \int_0^\infty G(t) dt.
\end{equation}
Furthermore, the storage modulus $G'(\omega)$ and loss modulus
$G''(\omega)$ were obtained from the stress relaxation modulus $G(t)$ as
follows:
\begin{align}
G'(\omega)  &= \omega \int_0^\infty G(t) \sin (\omega t) dt,\\
G''(\omega)  &= \omega \int_0^\infty G(t) \cos (\omega t) dt.
\end{align}
The magnitude of the complex viscosity, $\lvert\eta^*(\omega)\rvert$,
was calculated from the storage modulus $G'(\omega)$ and loss modulus
$G''(\omega)$ as follows:
\begin{equation}
\lvert\eta^*(\omega)\rvert
=
\frac{\sqrt{\left[G'(\omega)\right]^2+\left[G''(\omega)\right]^2}}{\omega}.
\end{equation}

In the NEMD simulations, uniform shear flow was generated using the
SLLOD equations of motion combined with Lees--Edwards boundary
conditions.~\cite{evans2008Statistical}
The flow and velocity-gradient directions were defined as the $x$ and
$y$ directions, respectively, and the time evolution of the $xy$
component of the pressure tensor, $P_{xy}$, was calculated.
The shear rate $\dot{\gamma}$ was varied from $10^{-5}$ to $10^{-1}$,
and the shear-rate-dependent viscosity $\eta(\dot{\gamma})$ was
calculated from the steady-state value of $P_{xy}$ as 
\begin{equation}
\eta(\dot{\gamma})  = -\frac{P_{xy}}{\dot{\gamma}}.
\end{equation}
Note that results at low shear rates were excluded when the $P_{xy}$
response could not be determined with sufficient statistical accuracy.

%%%%%%%%% Fig. 2 %%%%%%%%%%%%%%%%%%%%%%%%%%%%%%%%%%%%%%%%%%%%%%%%%%%%%%%%%%%%%%%%%%%%%%%%%%%%%%%%%%%%%%%%%%%%%%%%%%%%
\begin{figure*}[t]
\centering
 \includegraphics[width=\textwidth]{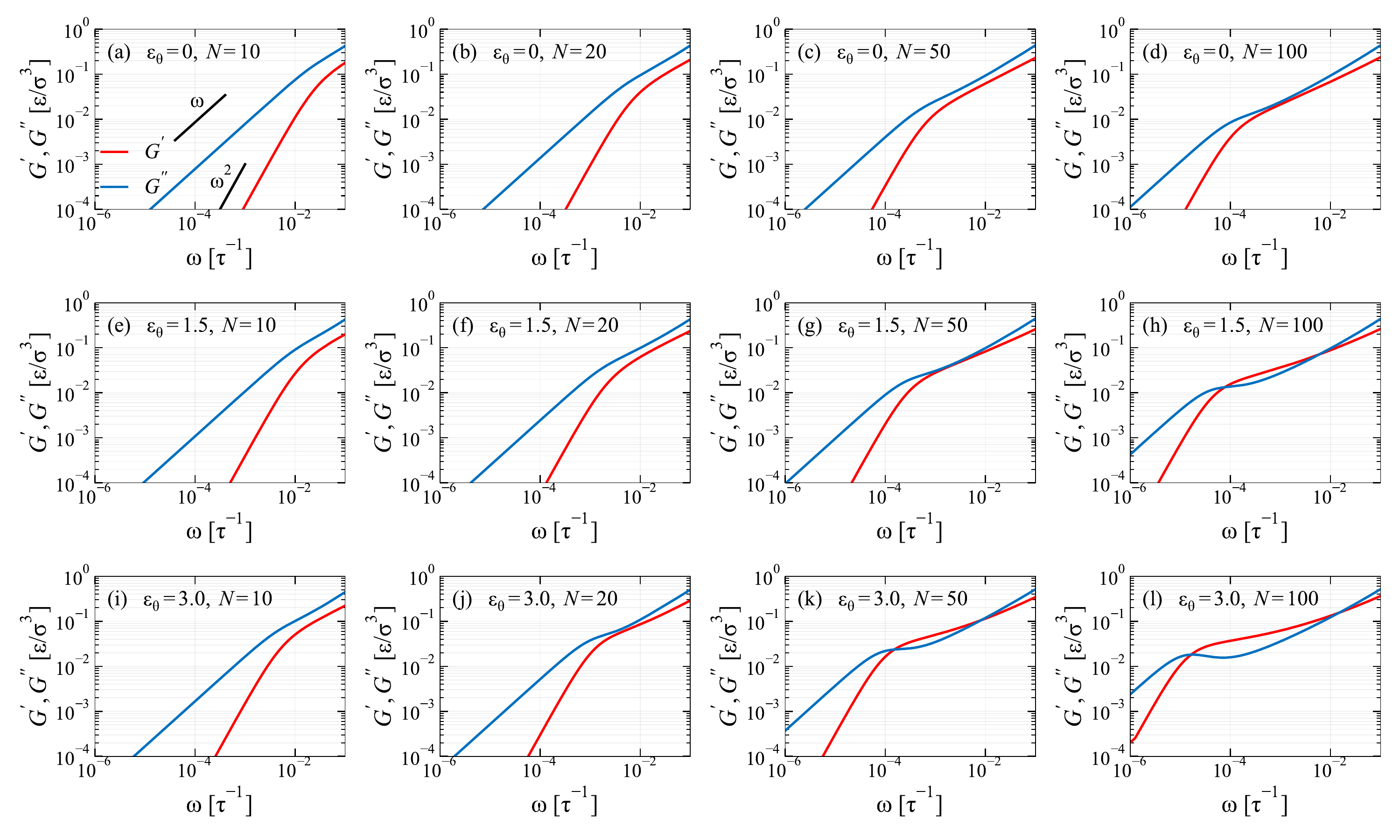}
\caption{Storage modulus $G'(\omega)$ (red) and loss modulus $G''(\omega)$ (blue)
 for linear polymers at different chain lengths $N$ and chain
 stiffnesses $\varepsilon_\theta$.
The chain lengths are $N=10$ [(a), (e), and (i)], $N=20$ [(b), (f), and
 (j)], $N=50$ [(c), (g), and (k)], and $N=100$ [(d), (h), and (l)]. The
 chain stiffnesses are $\varepsilon_\theta=0$ [(a)-(d)],
 $\varepsilon_\theta=1.5$ [(e)-(h)], and $\varepsilon_\theta=3.0$
 [(i)-(l)].
}
\label{fig:moduli_linear}
\end{figure*}
%%%%%%%%%%%%%%%%%%%%%%%%%%%%%%%%%%%%%%%%%%%%%%%%%%%%%%%%%%%%%%%%%%%%%%%%%%%%%%%%%%%%%%%%%%%%%%%%%%%%%%%%%%%%%%%%%%%%%

%%%%%%%%% Fig. 3 %%%%%%%%%%%%%%%%%%%%%%%%%%%%%%%%%%%%%%%%%%%%%%%%%%%%%%%%%%%%%%%%%%%%%%%%%%%%%%%%%%%%%%%%%%%%%%%%%%%%
\begin{figure*}[t]
\centering
 \includegraphics[width=\textwidth]{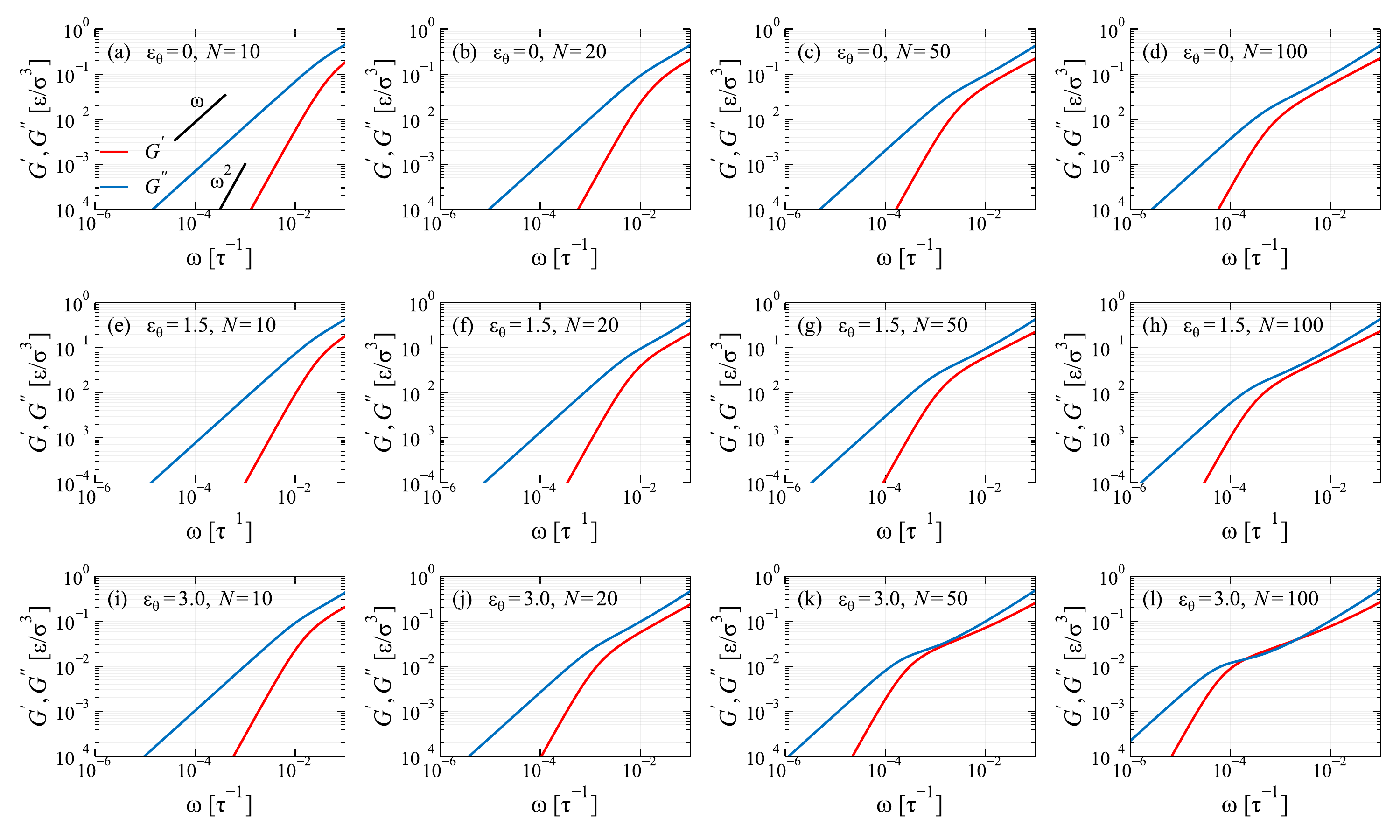}
\caption{Storage modulus $G'(\omega)$ (red) and loss modulus $G''(\omega)$ (blue)
 for ring polymers at different chain lengths $N$ and chain
 stiffnesses $\varepsilon_\theta$.
The chain lengths are $N=10$ [(a), (e), and (i)], $N=20$ [(b), (f), and
 (j)], $N=50$ [(c), (g), and (k)], and $N=100$ [(d), (h), and (l)]. The
 chain stiffnesses are $\varepsilon_\theta=0$ [(a)-(d)],
 $\varepsilon_\theta=1.5$ [(e)-(h)], and $\varepsilon_\theta=3.0$
 [(i)-(l)].
}
\label{fig:moduli_ring}
\end{figure*}
%%%%%%%%%%%%%%%%%%%%%%%%%%%%%%%%%%%%%%%%%%%%%%%%%%%%%%%%%%%%%%%%%%%%%%%%%%%%%%%%%%%%%%%%%%%%%%%%%%%%%%%%%%%%%%%%%%%%%

%%%%%%%%% Fig. 4 %%%%%%%%%%%%%%%%%%%%%%%%%%%%%%%%%%%%%%%%%%%%%%%%%%%%%%%%%%%%%%%%%%%%%%%%%%%%%%%%%%%%%%%%%%%%%%%%%%%%
\begin{figure*}[t]
\centering
 \includegraphics[width=0.8\textwidth]{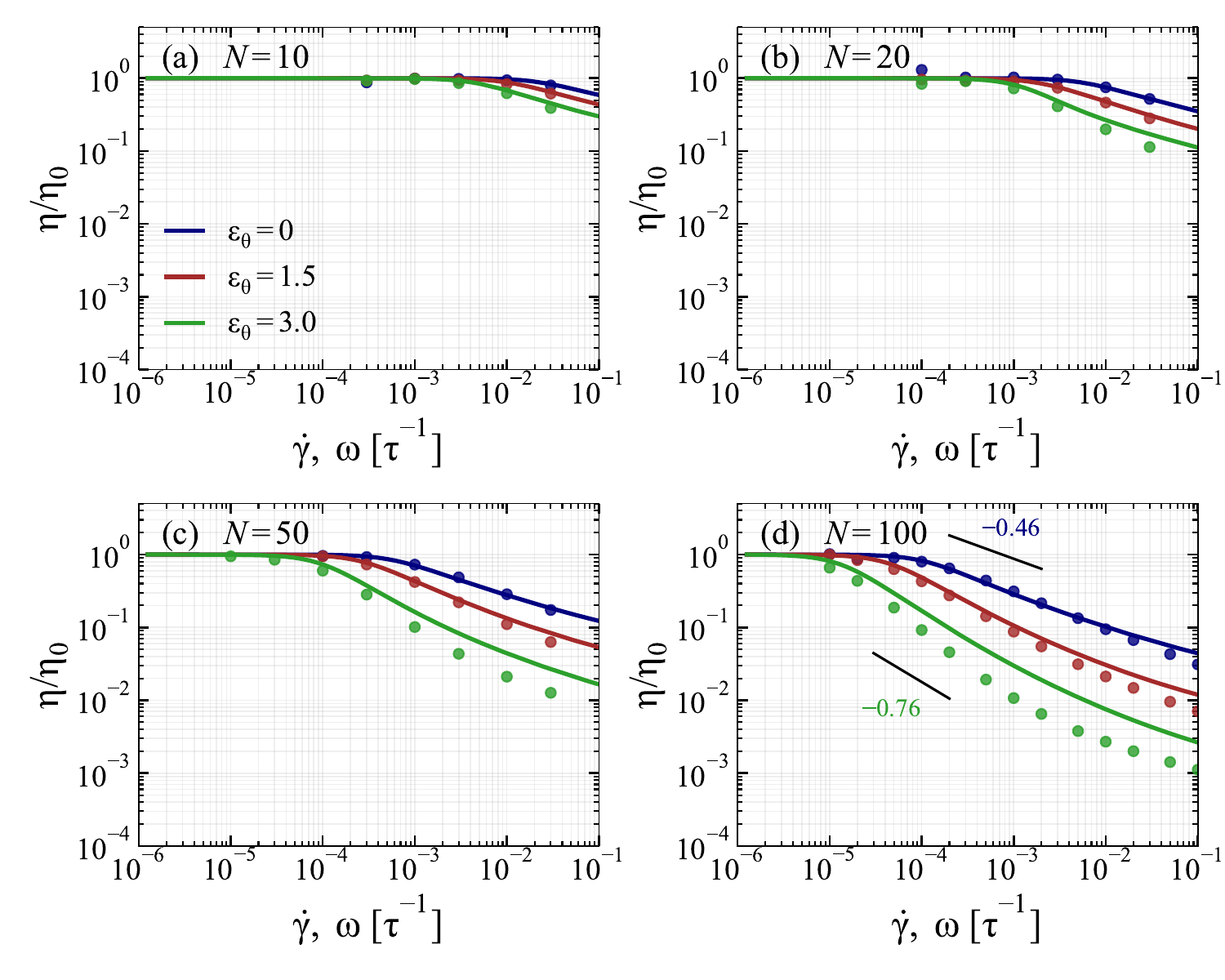}
\caption{Frequency dependence of the magnitude of the complex viscosity,
 $\lvert\eta^*(\omega)\rvert$ (solid lines), and shear-rate dependence
 of the steady-shear viscosity, $\eta(\dot{\gamma})$ (circles), for
 linear polymers. 
Both viscosities are normalized by the zero-shear viscosity $\eta_0$. 
The chain lengths are (a) $N=10$, (b) $N=20$, (c) $N=50$, and (d)
 $N=100$. 
The chain stiffnesses are $\varepsilon_\theta=0$ (blue), $1.5$ (red),
 and $3.0$ (green).
In panel (d), the short black lines indicate guides to the power-law
 behavior, $\eta(\dot{\gamma})\propto\dot{\gamma}^{-\alpha}$, in the
 shear-thinning regime for $N=100$, with the corresponding values of $\alpha$ for
 $\varepsilon_\theta=0$ and 3.0 shown in blue and green, respectively.
}
\label{fig:eta_linear}
\end{figure*}
%%%%%%%%%%%%%%%%%%%%%%%%%%%%%%%%%%%%%%%%%%%%%%%%%%%%%%%%%%%%%%%%%%%%%%%%%%%%%%%%%%%%%%%%%%%%%%%%%%%%%%%%%%%%%%%%%%%%%

%%%%%%%%% Fig. 5 %%%%%%%%%%%%%%%%%%%%%%%%%%%%%%%%%%%%%%%%%%%%%%%%%%%%%%%%%%%%%%%%%%%%%%%%%%%%%%%%%%%%%%%%%%%%%%%%%%%%
\begin{figure*}[t]
\centering
 \includegraphics[width=0.8\textwidth]{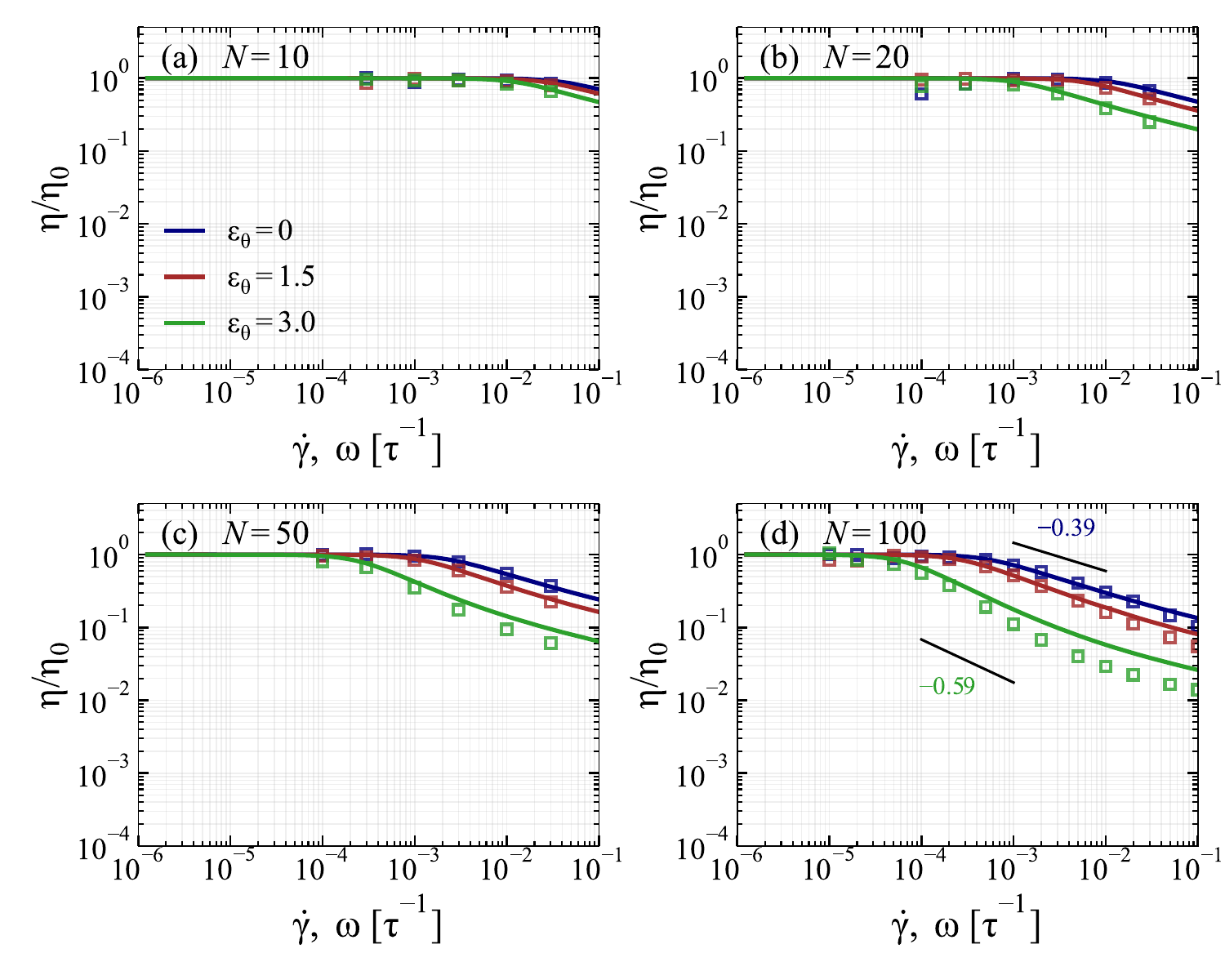}
\caption{Frequency dependence of the magnitude of the complex viscosity,
 $\lvert\eta^*(\omega)\rvert$ (solid lines), and shear-rate dependence
 of the steady-shear viscosity, $\eta(\dot{\gamma})$ (squares), for
 ring polymers. 
Both viscosities are normalized by the zero-shear viscosity $\eta_0$. 
The chain lengths are (a) $N=10$, (b) $N=20$, (c) $N=50$, and (d)
 $N=100$. 
The chain stiffnesses are $\varepsilon_\theta=0$ (blue), $1.5$ (red), and $3.0$ (green).
In panel (d), the short black lines indicate guides to the power-law
 behavior, $\eta(\dot{\gamma})\propto\dot{\gamma}^{-\alpha}$, in the
 shear-thinning regime for $N=100$, with the corresponding values of $\alpha$ for
 $\varepsilon_\theta=0$ and 3.0 shown in blue and green, respectively.
}
\label{fig:eta_ring}
\end{figure*}
%%%%%%%%%%%%%%%%%%%%%%%%%%%%%%%%%%%%%%%%%%%%%%%%%%%%%%%%%%%%%%%%%%%%%%%%%%%%%%%%%%%%%%%%%%%%%%%%%%%%%%%%%%%%%%%%%%%%%

\section{Results}

\subsection{Stress relaxation modulus and zero-shear viscosity}

Figure~\ref{fig:stress_relaxation}(a) shows the stress relaxation modulus
$G(t)$ for linear and ring polymers with $N=100$ at different chain
stiffnesses $\varepsilon_\theta$.
Although the corresponding results for the other chain lengths are not
shown, the qualitative features remain unchanged as the chain length
decreases, except that the relaxation time scale becomes shorter.
For $\varepsilon_\theta=0$ and $1.5$, the behavior of $G(t)$ is
consistent with that reported in previous
studies.~\cite{halverson2011Moleculara, behbahani2024Relaxation}
At short times, $G(t)$ is dominated by local segmental dynamics,
followed by a regime approximately described by the Rouse prediction,
$G(t)\propto t^{-1/2}$.
At longer times, the relaxation behavior depends strongly on molecular
architecture and chain stiffness.

For linear polymers, an intermediate-time plateau associated with
entanglement constraints becomes increasingly pronounced as
$\varepsilon_\theta$ increases. 
This chain-stiffness dependence of the stress relaxation behavior is
consistent with that reported in previous simulations of linear
Kremer--Grest polymer melts.~\cite{behbahani2026Stress} 
This behavior is
also consistent with the decrease in the entanglement length
$N_\mathrm{e}$ with increasing $\varepsilon_\theta$, as discussed in
Sec.~\ref{section_methods}.
At a fixed chain length, a smaller $N_{\mathrm e}$ corresponds to a
larger number of entanglements per chain and therefore stronger
constraints on chain relaxation.

In contrast to linear polymers, ring polymers exhibit faster stress
relaxation and do not show a comparably pronounced 
plateau regime.
This difference in $G(t)$ between linear and ring polymers is consistent
with that reported in a previous study.~\cite{parisi2021Nonlinear}
Although the dynamics of ring polymers are influenced by ring-specific topological
constraints, including inter-ring threadings, these constraints do not
produce the same relaxation behavior as conventional entanglements in
linear chains. 
Consequently, ring polymers can relax stress through conformational
rearrangements without the long-lived tube confinement characteristic of
entangled linear polymers, resulting in faster decay of $G(t)$. 
With increasing chain stiffness, however, a more pronounced plateau-like
regime emerges in $G(t)$ for ring polymers. 
This behavior may reflect stronger topological constraints associated
with increased interpenetration and threading between neighboring rings.

Figure~\ref{fig:stress_relaxation}(b) shows the dependence of the zero-shear viscosity,
$\eta_0$, on chain length $N$ 
for different chain stiffnesses $\varepsilon_\theta$.
For both linear and ring polymers, $\eta_0$ increases with increasing
chain length and chain stiffness. 
At the same $N$ and $\varepsilon_\theta$, linear polymers exhibit a
higher $\eta_0$ than ring polymers. 
These trends are consistent with the differences in the stress
relaxation behavior shown in Fig.~\ref{fig:stress_relaxation}, since
$\eta_0$ is determined by the time integral of $G(t)$.

\subsection{Storage and loss moduli}

Figures~\ref{fig:moduli_linear} and \ref{fig:moduli_ring} show the
angular-frequency dependence of the storage modulus, $G'(\omega)$, and
loss modulus, $G''(\omega)$, for linear and ring polymers, respectively,
at different chain lengths $N$ and chain stiffnesses
$\varepsilon_\theta$.
In the low-frequency regime, $G''(\omega)$ exceeds $G'(\omega)$,
indicating predominantly viscous behavior.
Both moduli approach the terminal-flow behavior,
$G'(\omega)\propto\omega^2$ and $G''(\omega)\propto\omega$.
For linear polymers with $\varepsilon_\theta=0$, the frequency
dependence of $G'(\omega)$ and $G''(\omega)$ is consistent with that
reported in previous molecular dynamics studies of fully flexible
Kremer--Grest polymer
melts.~\cite{likhtman2007Linear, adeyemi2022Equilibrium, behbahani2024Relaxation}
As the chain length and chain stiffness increase, the elastic response
becomes more pronounced, and an intermediate-frequency regime with
$G'(\omega)>G''(\omega)$ emerges. 
For sufficiently long and stiff
chains, $G'(\omega)$ and $G''(\omega)$ exhibit two distinct crossover
frequencies, defining an intermediate-frequency window in which the
elastic response dominates the viscous response. 
This window broadens
with increasing chain length and chain stiffness, consistent with the
increasingly pronounced plateau in $G(t)$.

For ring polymers, as shown in Fig.~\ref{fig:moduli_ring}, both
$G'(\omega)$ and $G''(\omega)$ are generally smaller than those for
linear polymers.
For short chains, $G''(\omega)$ exceeds $G'(\omega)$ over the entire
frequency range examined, irrespective of chain stiffness.
As the chain length increases, an intermediate-frequency regime with
$G'(\omega)>G''(\omega)$ can emerge for higher chain stiffness.
However, even for longer and stiffer rings, the two crossover
frequencies are less clearly separated than in linear polymers, and a
well-defined elastic-dominated window is not as apparent.
This weaker separation between viscous- and elastic-dominated regimes
is consistent with the less pronounced plateau-like behavior in
$G(t)$ for ring polymers.

\subsection{The Cox--Merz rule}

Finally, we compare the complex viscosity obtained from equilibrium
simulations with the steady-shear viscosity obtained from NEMD
simulations to examine the validity of the Cox--Merz rule. 
Figures~\ref{fig:eta_linear} and \ref{fig:eta_ring} show
$\lvert\eta^*(\omega)\rvert$ as a function of angular frequency $\omega$
and $\eta(\dot{\gamma})$ as a function of shear rate $\dot{\gamma}$ for
linear and ring polymers, respectively, at different chain lengths $N$
and chain stiffnesses $\varepsilon_\theta$. 
Both viscosities are normalized by the corresponding
zero-shear viscosity $\eta_0$ shown in
Fig.~\ref{fig:stress_relaxation}(b) to highlight their shear-thinning
behavior.
Parisi \textit{et al.} reported the shear-rate dependence of
$\eta(\dot{\gamma})$ for linear and ring polymers at
$\varepsilon_\theta=1.5$ for several chain lengths.~\cite{parisi2021Nonlinear}
For $N=100$ and $\varepsilon_\theta=1.5$, our results are in good
agreement with their simulation results.

We first consider the steady-shear viscosity. 
In the low-shear-rate regime, $\eta(\dot{\gamma})$ increases with
increasing chain stiffness. 
At higher shear rates, however, the viscosity decreases more strongly
for stiffer chains, and this effect becomes particularly pronounced at
larger $N$. 
For short chains, such as $N=10$, the shear-rate dependence is
relatively weak over the range examined. 
For the same $N$ and $\varepsilon_\theta$, linear polymers exhibit a
higher viscosity than ring polymers in the low-shear-rate regime. 
In addition, the decrease in viscosity with increasing shear rate is
generally more pronounced for linear polymers. These
architecture-dependent differences are consistent with the distinct
relaxation behavior of linear and ring polymers discussed above.
As shown in Figs.~\ref{fig:eta_linear}(d) and \ref{fig:eta_ring}(d), the
shear-thinning behavior can be further characterized by the power-law
relation $\eta(\dot{\gamma})\propto\dot{\gamma}^{-\alpha}$. 
For $N=100$, the exponent $\alpha$ increases from 0.46 to 0.76 with
increasing chain stiffness from $\varepsilon_\theta=0$ to 3.0 for linear
polymers, whereas it increases from 0.39 to 0.59 for ring polymers. 
Thus, increasing chain stiffness enhances shear thinning for both
architectures, with the effect being more pronounced for linear
polymers.

We next compare $\lvert\eta^*(\omega)\rvert$ and $\eta(\dot{\gamma})$ at
corresponding values of $\omega$ and $\dot{\gamma}$. 
For fully flexible chains with $\varepsilon_\theta=0$, the two
viscosities show good agreement, indicating that the Cox--Merz rule
approximately holds over the range examined. 
In contrast, as the chain stiffness increases, deviations between
$\lvert\eta^*(\omega)\rvert$ and $\eta(\dot{\gamma})$ become
increasingly apparent, particularly for linear polymers. 
At $\varepsilon_\theta=3.0$, $\lvert\eta^*(\omega)\rvert$ exceeds
$\eta(\dot{\gamma})$ at high corresponding values of $\omega$ and
$\dot{\gamma}$, and the deviation becomes more pronounced with
increasing chain length. 
For short chains such as $N=10$, by contrast, the two viscosities remain
in relatively good agreement irrespective of chain stiffness.

The architecture dependence is particularly evident for longer and
stiffer chains. 
For linear polymers, the discrepancy between
$\lvert\eta^*(\omega)\rvert$ and $\eta(\dot{\gamma})$ becomes
substantial with increasing $N$ and $\varepsilon_\theta$, whereas for
ring polymers the discrepancy remains relatively small even at larger
$N$ and higher $\varepsilon_\theta$. 
Thus, the breakdown of the Cox--Merz rule is much more pronounced for
long and stiff linear polymers than for the corresponding ring polymers. 
It should be noted, however, that Parisi \textit{et
al.}~\cite{parisi2021Nonlinear} reported a pronounced deviation from the
Cox--Merz rule for much longer ring polymers, extending up to $N=800$ at
$\varepsilon_\theta = 1.5$,
indicating that the relatively good agreement observed here for ring
polymers is limited to the chain-length range examined in the present
study.
This difference suggests that the validity of the Cox--Merz
correspondence is closely associated with the different responses of the
two molecular architectures to steady shear.

This difference can be understood in relation to the
architecture-dependent conformational response to strong shear flow
observed in our recent study.~\cite{sakamaki2026Elucidationa}
We found that linear polymers undergo greater flow-induced extension
than ring polymers, as reflected in the $xx$ component of the gyration
tensor, whereas the $yy$ component, which characterizes the molecular
dimension in the velocity-gradient direction, is closely correlated
with the shear viscosity.
This correlation between viscosity and chain conformation under
shear flow 
is also consistent with several recent studies.~\cite{xu2014Shear,
xu2017Probing, nikoubashman2017Equilibrium,
gurel2023Shear, uneyama2025Radius}
For linear polymers, the presence of chain ends allows substantial
flow-induced extension and alignment, which may lead to a partial
release of entanglement constraints under strong shear.
In contrast, ring polymers have no chain ends and retain rotational
degrees of freedom even under flow-induced deformation, resulting in
smaller conformational changes than in linear polymers.
These results suggest that, as the flow-induced conformational changes
become more pronounced, the steady-shear response increasingly
deviates from the relaxation behavior characterized by linear
viscoelasticity near equilibrium. Consequently, the complex viscosity
$\lvert\eta^*(\omega)\rvert$ can no longer accurately predict the
steady-shear viscosity $\eta(\dot{\gamma})$, leading to a pronounced
breakdown of the Cox--Merz rule.

\section{Conclusions}

In this study, we systematically varied the chain length and chain
stiffness of coarse-grained polymer melts described by the
Kremer--Grest model for two molecular architectures, linear and ring
polymers, to identify the conditions under which the Cox--Merz rule
holds and the factors associated with its breakdown. Our results show
that the validity of the Cox--Merz rule depends strongly on chain
length, chain stiffness, and molecular architecture. 
For short chains and, more generally, for flexible chains,
the magnitude of the complex viscosity,
$\lvert\eta^*(\omega)\rvert$, agrees well with the steady-shear
viscosity, $\eta(\dot{\gamma})$, at corresponding values of $\omega$
and $\dot{\gamma}$. For linear polymers, however, the discrepancy
between the two viscosities becomes increasingly pronounced with
increasing chain length and chain stiffness, 
whereas for ring polymers the discrepancy remains relatively small even
for longer and stiffer chains within the range examined here.
The increasing deviation from the Cox--Merz rule with increasing chain
length is consistent with the chain-length dependence reported by
Parisi \textit{et al.}~\cite{parisi2021Nonlinear} 
Our results further show that
this tendency is strongly enhanced by increasing chain stiffness,
particularly for linear polymers.

For long and stiff linear polymers, entanglement effects are
pronounced, and strong steady shear induces substantial chain extension
and alignment. Such flow-induced conformational changes can modify the
entanglement constraints that govern relaxation near equilibrium,
leading to a reduction in $\eta(\dot{\gamma})$ relative to
$\lvert\eta^*(\omega)\rvert$. In contrast, ring polymers exhibit
smaller flow-induced conformational changes and are subject to
architecture-specific topological constraints that differ from the
conventional entanglement constraints of linear polymers. The weaker
breakdown of the Cox--Merz rule for ring polymers is consistent with
this distinct response to steady shear.

These results demonstrate that the applicability of the Cox--Merz rule
is closely associated with the extent to which steady shear alters the
molecular conformations and topological constraints underlying
equilibrium relaxation. When these changes become substantial, as in
long and stiff linear polymers under strong shear, the steady-shear
response deviates markedly from the relaxation behavior characterized
by linear viscoelasticity near equilibrium. Consequently, the complex
viscosity $\lvert\eta^*(\omega)\rvert$ can no longer accurately predict
the steady-shear viscosity $\eta(\dot{\gamma})$, resulting in a
pronounced breakdown of the Cox--Merz rule.

\begin{acknowledgments}
This work was supported by 
JSPS KAKENHI
Grant Nos.~\mbox{JP25K00968} and \mbox{JP23H02622}.
We acknowledge support from the Fugaku Supercomputing Project
 (Nos.~\mbox{JPMXP1020230325} and \mbox{JPMXP1020230327}) and 
the Data-Driven Material Research Project (No.~\mbox{JPMXP1122714694})
from the
Ministry of Education, Culture, Sports, Science, and Technology.
The numerical calculations were performed at Research Center of
Computational Science, Okazaki Research Facilities, National Institutes
of Natural Sciences (Project: 26-IMS-C051).
\end{acknowledgments}

\section*{AUTHOR DECLARATIONS}

\section*{Conflict of Interest}
The authors have no conflicts to disclose.

\section*{Data availability statement}

Data that support the findings of this study are available from
the corresponding author upon request.

%aipnum4-2.bst 2019-01-14 (MD) hand-edited version of apsrev4-1.bst

\end{document}